# Theoretical study of asymmetric A-π-D-π-D-π-A' tribranched organic sensitizer for Dye-sensitized solar cells

**Geon Hyeong Lee[1] and Young Sik Kim[1,2*]**

[1] *Department of Information Display, Hongik University, Seoul 121-791*

[2] *Department of Science, Hongik University, Seoul 121-791*

An asymmetric A-π-D-π-D-π-A' tribranched organic dye (**dye1**) with a cyanoacrylic acid and an indolinum carboxyl acid as electron acceptors and a triphenylamine as an electron donor was designed and theoretically investigated for dye-sensitized solar cells (DSSCs). **Dye1** was compared to reference well-known dyes with single electron acceptors (**D5** and **JYL-SQ6**). Density functional theory and time-dependent density functional theory calculations were used to estimate the photovoltaic properties of the dyes. Due to the different lowest unoccupied molecular orbital levels of each acceptor and the energy antenna of the dual electron donor (D-π-D), the absorption spectrum of each branch displayed different shapes. Considering the overall properties, the asymmetric A-π-D-π-D-π-A' tribranched organic dye exhibited high conversion efficiency performance for DSSCs. The findings of this work suggest that optimizing the branch of electron donors and acceptors in dye sensitizers based on asymmetric A-π-D-π-D-π-A' tribranched organic dye produces good photovoltaic properties for DSSCs.

Email: youngkim@hongik.ac.kr

Fax: +82-2-3142-0335

## I. INTRODUCTION

Much attention has recently been focused on using dye-sensitized solar cells (DSSCs) as possible low-cost photovoltaic devices [1]. The performance and stability of DSSC devices have been studied and developed significantly over the past decade [2-4]. Among the components of DSSCs, the sensitizer is a crucial element that significantly influences the power conversion efficiency and stability of the device. Currently, perovskite dyes offer the greatest efficiency among DSSCs (close to 15%). Zinc porphyrin dye also offers great efficiency for DSSCs (over 12%) in conjunction with the Co(II/III)tris(bipyridyl)-based redox electrolyte, and ruthenium polypyridyl dyes offers good efficiency for DSSCs (over 11%) in combination with a voltaic iodide/triiodide mixture as the electrolyte [5-7]. However, metal-based dyes, such as ruthenium or zinc complexes, are not suitable in terms of cost efficiency or environmental friendliness [8-9]. Thus, metal-free organic dyes have received attention as alternatives to metal-based dyes, due to their high molar extinction coefficient, simple synthesis and purification, low cost, and environmental friendliness.

Numerous metal-free organic dyes for DSSCs, such as coumarin-,[10] merocyanine-,[11] indoline-,[12] porphyrin-,[13] and fluorine-[14] based organic dyes, have been developed, and all exhibit good DSSC performance. A donor-π conjugated linker-acceptor (D-π-A) structure is generally used for metal-free organic dyes [15]; the dual-donor or dual-acceptor group of dyes can increase the photovoltaic performance of these dyes [16-17]. Previously, we designed several organic sensitizers featuring dual heteroleptic acceptors, showing that the absorption spectrum of the dual heteroleptic acceptor dye represents a mixed feature of two types of acceptor dyes. We found that DSSCs based on heteroleptic dual electron acceptors exhibited better photovoltaic performance than dual homoleptic acceptors [18-19]. Dual donor-based dyes can also increase light-harvesting ability through an intramolecular energy transfer from an antenna group. The antenna group of the hole-transport materials can also delay the charge-recombination process through the physical separation of the holes, moving them spatially away from the electrons in the semiconductor, thus improving the efficiency of the DSSCs [20-21].

In this work, we investigated the effect of an asymmetric A-π-D-π-D-π-A' tribranched organic dye (**dye1**) with a cyanoacrylic acid and an indolinum carboxyl acid as electron acceptors and a triphenylamine (TPA) and an additional TPA as an antenna group of the dye to transfer the energy to the basic dye as a photosensitizer for DSSCs. The properties of **dye1** were compared with those of well-known 3-(5-(4-(diphenyl-amino) styryl) thiophen-2-yl)-2-cyanoacrylic acid (**D5**) and 5-carboxy-2-[[3-(5-(5-(4-(bis(4-(hexyloxy) phenyl)amino)phenyl)thiophen-2-yl)thiophen-2-yl)-2-hydroxy-4-oxo-2-cyclobuten-1-lidene] methyl]-3,3-dimethyl-1-butyl-3H-indolium (**JYL-SQ6**) photosensitizers. We investigated the charge transfer and absorption mechanism of the dyes to understand the role of a dye sensitizer. This theoretical study provides insight for developing more efficient dyes for DSSCs.

## II. EXPERIMENTS AND DISCUSSION

### 1. Computational Methods

To gain insight into the factors responsible for the absorption spectral response and the conversion efficiency, we perform DFT and TD-DFT calculations on the ground state of the dyes. This computational procedure allows us to provide a detailed assignment of the excited states involved in the absorption process. The geometries and energy levels of the molecular orbital were calculated using the DFT method, and the absorption spectrum was calculated at optimized ground state geometries by the TD-DFT method. Although, the molecular orbital's energy levels can predict the trend of an energy gap in the absorption spectrum, such data do not exactly match in the absorption spectrum. Since the absorption process has time dependency, the absorption spectrum in this study originates from TD-DFT.

The geometries in the gas phase were optimized by the DFT method using the B3LYP/6-31G(d) in the Gaussian 09 program package. Electronic populations of the highest occupied molecular orbital (HOMO) and the lowest unoccupied molecular orbital (LUMO) were calculated to show the position of the localization of electron populations along with the calculated molecular orbital energy diagram.

TD-DFT calculations were made using the same basis set and two kinds of hybrid Functional MPW1K and B3LYP at the ground state optimized geometries by means of the C-PCM algorithm. The absorption spectrum and oscillator strengths were calculated at optimized ground state geometries for the lowest 20 singlet-singlet excitations up to a wavelength of 350 nm. The simulation of the absorption spectra was performed by a Gaussian convolution with a full width at half-maximum (fwhm) = 0.3eV.

## 2. Results and Discussion

The schematic molecular structures of the TPA-based organic dyes considered in this study are shown in Fig. 1. **Dye1** has a common core TPA structure, with cyanoacrylic acid and indolinum as a heteroleptic electron dual acceptor. In order to understand the electronic transitions of **dye1**, TD-DFT calculations on absorption spectra in ethanol were performed with two types of hybrid functional MPW1K and B3LYP, based on the optimized geometries.

Table 1 presents a comparison between the calculated maximum absorption wavelength ($\lambda_{max}$) using different functionals and the experimental $\lambda_{max}$ (2.53eV/ 490nm) for **D5** and (1.91eV/ 650nm) for **JYL-SQ6** as a reference material [22]. The B3LYP functional undervalues the $\lambda_{max}$ values. The $\lambda_{max}$ and relative transition probabilities from MPW1K [(2.55eV/ 485nm) for **D5** and (1.96eV/ 632nm) for **JYL-SQ6**] agree best with the experimental data. Therefore, we performed the TD-DFT calculation with the hybrid functional MPW1K to determine the absorption spectra.

The calculated absorption energy, oscillator strength, and major composition in terms of the molecular orbital contributions to the dyes are shown in Table 1. Only excitations above 350 nm in absorption energy were considered in the table. The absorption peaks were dominated by two type transitions: intramolecular charge-transfer (ICT) and π→π* transitions. As shown in Table 1, the absorption peaks in the long wavelength region were dominated by ICT, and the absorption peak in the short wavelength region was dominated by π→π* transitions. ICT takes place between donor and acceptor and π→π* transitions take place inside TPA-π-TPA, which can be considered as an antenna effect.

The ultraviolet–visible (UV-vis) absorption spectra of the TPA-based dyes are shown in Fig. 2. The lowest absorption peaks of **D5** and **JYL-SQ6** were 485 nm and 632 nm, respectively, which were ascribed to HOMO→ LUMO transitions. **Dye1** shows three absorption bands. The major absorption peaks of the heteroleptic electron acceptor of **dye1** were around 636 nm and 499 nm, which were ascribed to HOMO → LUMO and HOMO-1 → LUMO+1 transitions, respectively. The peak around 381 nm was attributed to HOMO → LUMO+2 (see Table 1). The overall absorption bands of **dye1** had stronger intensities than those of other dyes. Thus, it is expected that **dye1** would exhibit the best absorbance among the dyes in this study.

The schematic of the calculated molecular orbital energy levels of **dye1** are shown in Fig. 3. In the device, $TiO_2$ was used as a semiconductor with a conduction band (CB) of -3.44 eV [23], and all the LUMO levels of the dyes were located above the $TiO_2$ LUMO level. Therefore, all of the dyes had sufficient driving force for electron injection into $TiO_2$. This result indicated that the energy band gap of **dye1** was decreased by the heteroleptic electron acceptor group. The LUMO energy level of **dye1** decreased compared to those of **D5** and **JYL-SQ6** because the LUMO of **dye1** interacted with the additional LUMO+1 due to the **D5** moiety. This explains why the LUMO level of **dye1**, with dual heteroleptic electron acceptors, decreased more than those of **D5** and **JYL-SQ6**, with single electron acceptor groups. Thus, the reduction in the energy band gap of **dye1** is represented as a red-shifted energy band in the absorption spectrum [11].

The electron distributions of some frontier molecular orbitals for **dye1** are shown in Fig. 4.

Molecular orbital analysis confirmed that the HOMO and HOMO-1 of **dye1** were mainly delocalized over the whole electron donor part with TPA. The LUMO was mainly localized over the indolinum anchoring part. LUMO+1 was localized over the cyanoacrylic moiety attached to the phenyl linker. By introducing the TPA-π-TPA antenna group, LUMO+2 and LUMO+3 were localized over the TPA-π-TPA antenna. This implied that LUMO+2 and LUNO+3 originated from the TPA-π-TPA antenna, contributing to the large π→π* absorbance around 381nm, as shown in Fig. 2.

In a study on the intramolecular energy transfer ($E_nT$) and charge transfer (ICT) of TPA-based dye, the antenna group of the dye transferred the energy to the basic dye. This $E_nT$ process took place through Förster energy transfer. For effective intramolecular $E_nT$, the emission band of the antenna group should substantially overlap the absorption band of the basic dye. Figure 5 shows a schematic drawing of the intramolecular $E_nT$ process of the dyes.

We confirmed that **dye1** showed broader and more red-shifted absorption spectra and a higher molar extinction coefficient than the basic dye did. This demonstrates that the emission band of the antenna group and the absorption band of the basic dyes should be considered for effective intramolecular $E_nT$ and photovoltaic properties of the basic dyes when adding an electron donor as an antenna to a single donor-based dye for effective DSSCs.

Absorption properties as a photovoltaic device can be provided directly in terms of incident photon-to-current conversion efficiency (IPCE) or light-harvesting efficiency (LHE). IPCE and LHE can be expressed as follows:

$$\text{IPCE}(\lambda) = \text{LHE}(\lambda) \cdot \Phi_{inj} \cdot \eta_{reg} \cdot \eta_{cc}, \qquad (1)$$

$$\text{LHE} = 1 - 10^{-A} = 1 - 10^{-f}, \qquad (2)$$

where $\Phi_{inj}$, $\eta_{reg}$ and $\eta_{cc}$ are the quantum yield of charge injection and the dye regeneration and charge collection efficiencies, respectively. Although A is the absorbance of the film, it can be represented by the oscillator strength (f) in the calculated absorption spectrum. If $\Phi_{inj}$, $\eta_{reg}$ and $\eta_{cc}$ are all assumed to be 100% for the dyes, it can be expressed as IPCE ≈ LHE. Table 1 shows the calculated LHEs obtained from the UV-vis absorption spectra of the dyes. The LHEs of **dye1** were better than those of the **D5** and **JYL-SQ6** basic dyes. Thus, we suggest that **dye1** would have better photovoltaic efficiency that the other dyes.

## III. CONCLUSION

Asymmetric A-π-D-π-D-π-A' tribranched **dye1**, with a cyanoacrylic acid and an indolinum carboxyl acid as electron acceptors and TPA as an electron donor, was designed and investigated for its electronic and optical properties as a DSSC device. To obtain a better understanding of its role as a dye sensitizer, particularly of its electronic structure and excited-state properties, we performed DFT and TD-DFT calculations. For these dye sensitizers, the heteroleptic dual electron acceptor and both cyanoacrylic and indolinum moieties could be used fully for the photovoltaic properties of **dye1**. In addition, we confirmed that adding an antenna group provided better photovoltaic properties. This study suggests that **dye1**, with a heteroleptic electron acceptor and the addition of a TPA-π-TPA antenna group, displayed the highest performance among the dyes in this study in terms of conversion efficiency for DSSCs.

## ACKNOWLEDGEMENT

This research was supported by the Basic Science Research Program through the National Research Foundation of Korea (NRF) funded by the Ministry of Education, Science and Technology (2010-0021668).

Table. 1. Calculated TD-DFT excitation energies (eV, nm), oscillator strengths (f), light-harvesting efficiency (LHE), and composition in terms of molecular orbital contributions of D5, JYL-SQ6 and dye1.

| Dye | E (eV, nm) | f | LHE | Major composition | Transition character |
|---|---|---|---|---|---|
| D5 | 2.55(485.0)<br>2.53(490.0)* | 1.799 | 0.984 | HOMO->LUMO (90%) | ICT + π→π* |
| JYL-SQ6 | 1.96(632.1)<br>1.91(650.0)* | 2.580 | 0.997 | HOMO->LUMO (90%) | ICT + π→π* |
| Dye1 | 1.95(635.9) | 2.663 | 0.998 | H-1->LUMO (32%), HOMO->LUMO (59%) | ICT + π→π* |
| | 2.49(498.6) | 1.956 | 0.989 | H-1->L+1 (40%), HOMO->L+1 (34%) | ICT + π→π* |
| | 3.25(381.0) | 1.026 | 0.906 | H-1->L+3 (10%), HOMO->L+2 (59%), HOMO->L+3 (11%) | π→π* |

*: Experimental data on ref. 22

Figure Captions

Fig. 1. Molecular structures of the dyes: (a) **D5**, (b) **JYL-SQ6**, and (c) **dye1**

Fig. 2. Calculated time-dependent density functional theory ultraviolet–visible absorption spectra of **D5**, **JYL-SQ6**, and **dye1**

Fig. 3. Schematic molecular orbital energy levels of **D5**, **JYL-SQ6**, and **dye1**

Fig. 4. Electron distributions of molecular orbitals of **dye1**

Fig. 5. Schematic drawing of the intramolecular $E_nT$ process of **dye1**

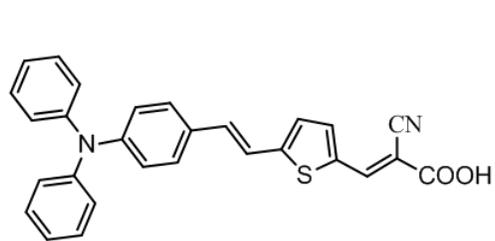
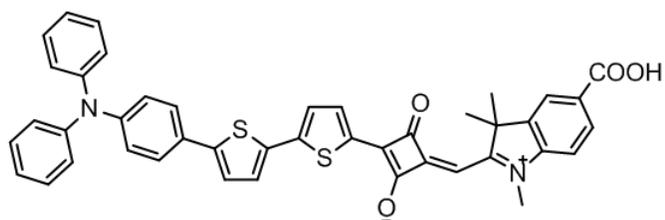
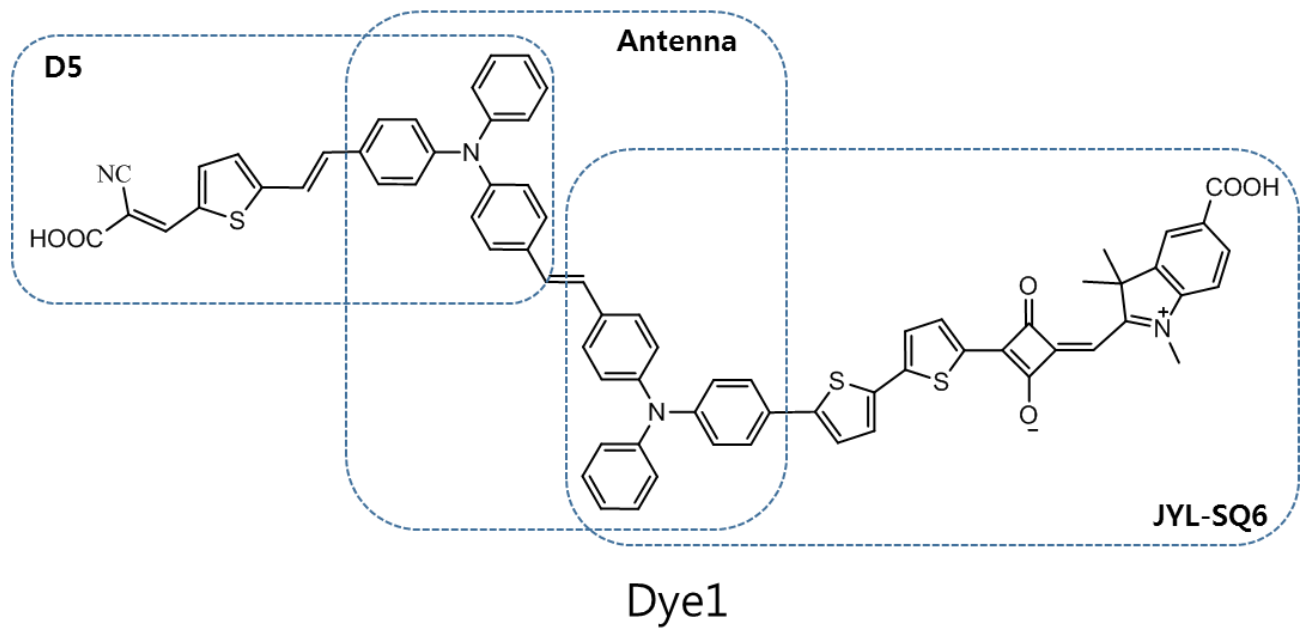

Fig. 1

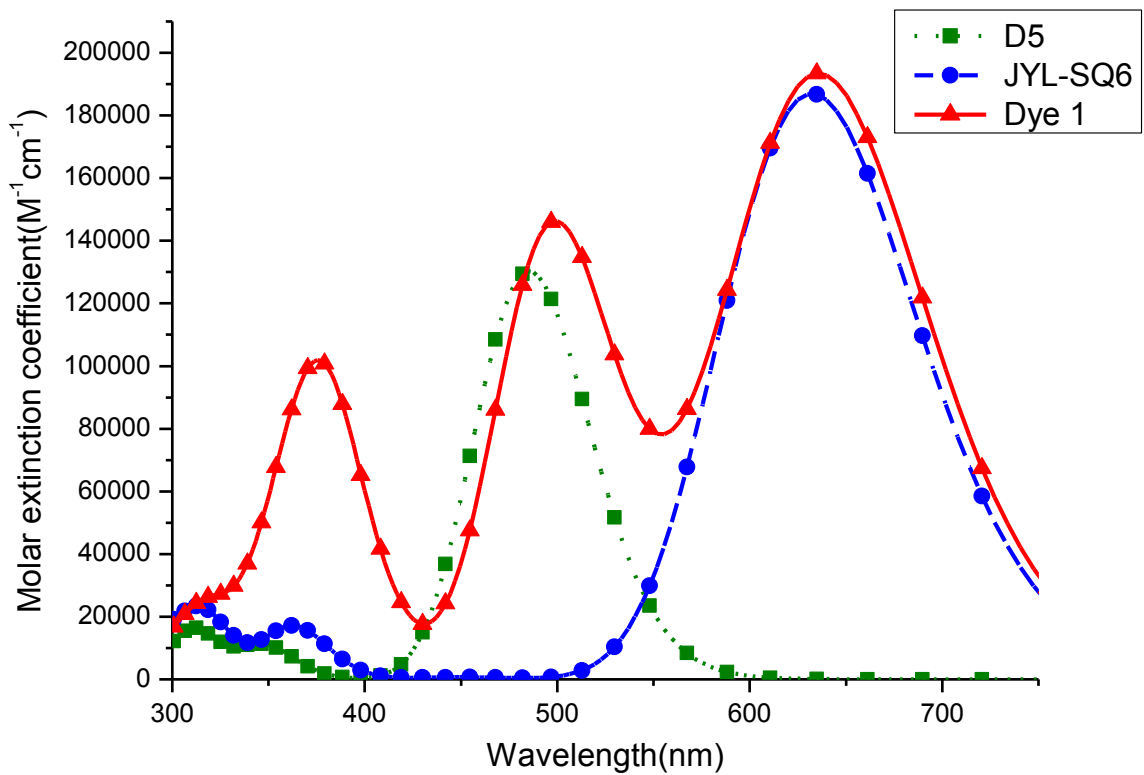

Fig. 2

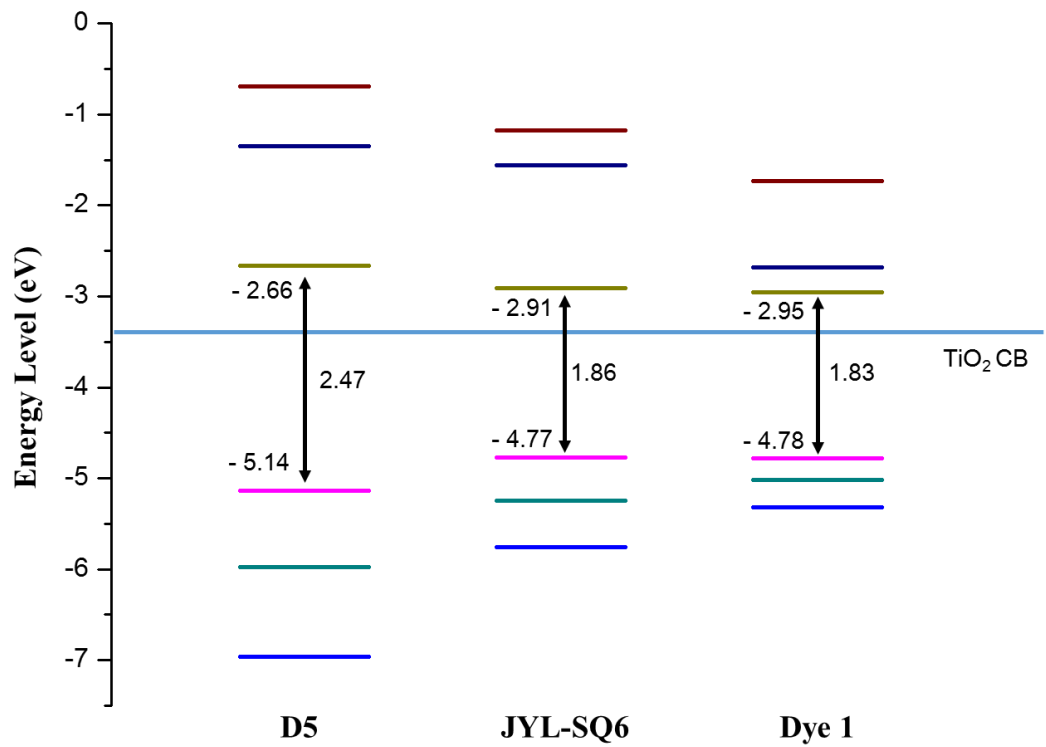

Fig. 3

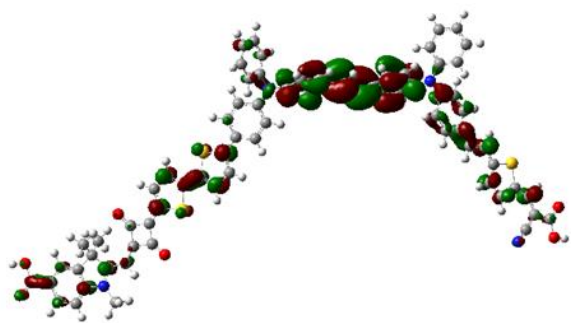

**LUMO+3**

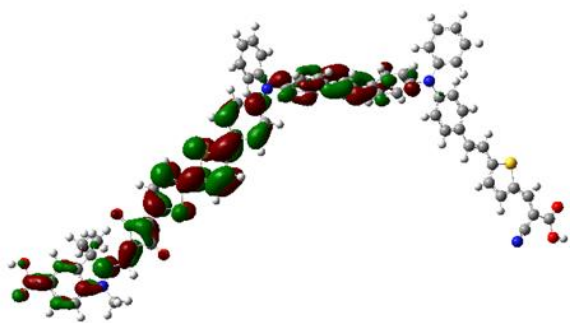

**LUMO+2**

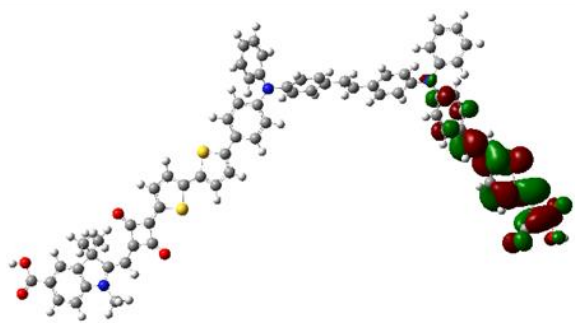

**LUMO+1**

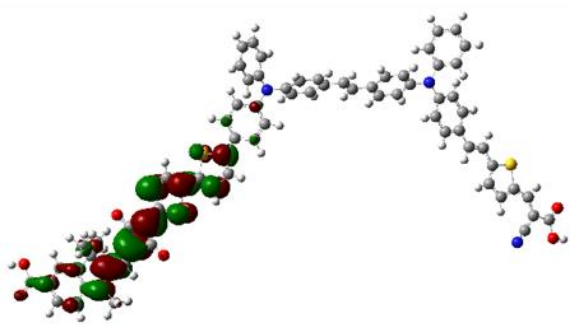

**LUMO**

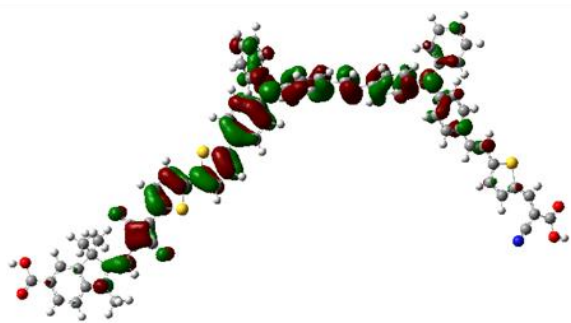

**HOMO**

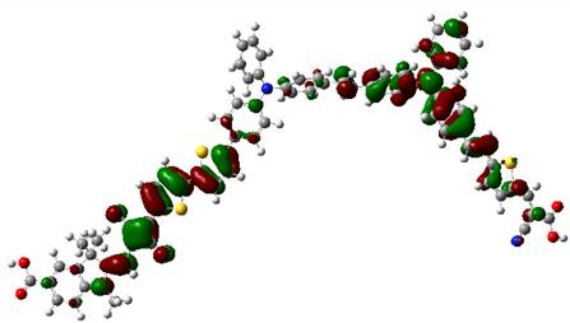

**HOMO-1**

Fig. 4

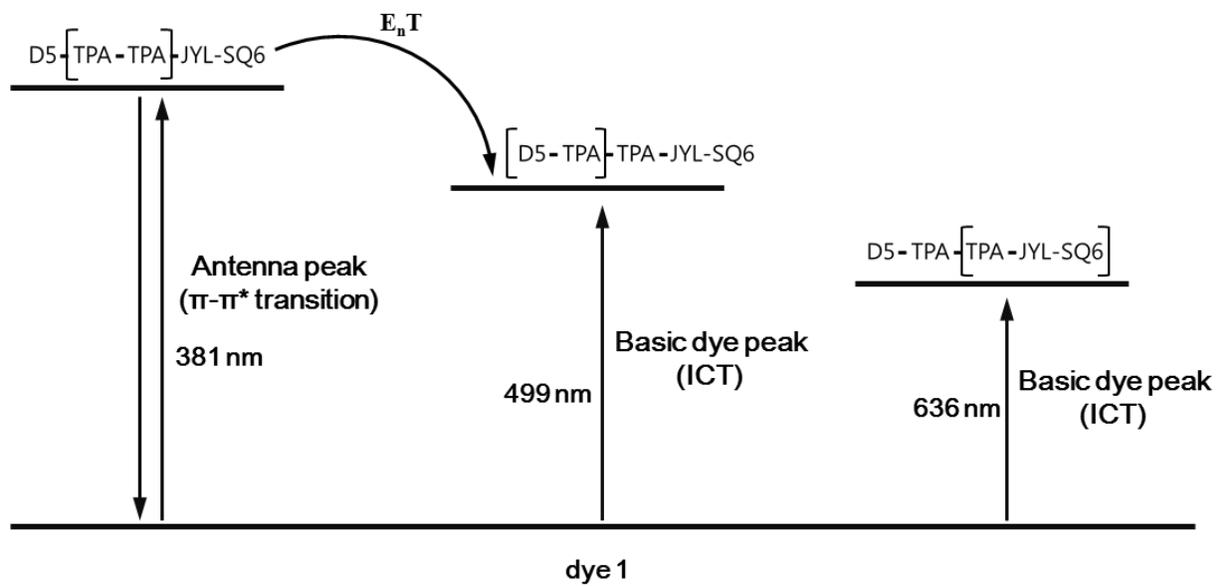

Fig. 5